\documentclass[
aps,
 amsmath,amssymb,
reprint,
superscriptaddress,
floatfix,
]{revtex4-1}

\usepackage{graphicx}
\usepackage{dcolumn}
\usepackage{bm}

\usepackage{xcolor}
\usepackage{booktabs}
\usepackage{wasysym}
\usepackage[uncertainty-mode = separate]{siunitx}
\usepackage[utf8]{inputenc}
\usepackage[T1]{fontenc}
\usepackage{mathptmx}
\usepackage{etoolbox}
\usepackage{hyperref}
\usepackage{stmaryrd}

\hypersetup{
    colorlinks = true,
    linkcolor = blue,
    citecolor = blue,
    urlcolor=blue,
}

\newcommand{\blue}[1]{\textcolor{blue}{#1}} 

\usepackage{tabularx}
\usepackage{appendix}
\usepackage{caption}
\makeatletter
\def\@email#1#2{%
 \endgroup
 \patchcmd{\titleblock@produce}
  {\frontmatter@RRAPformat}
  {\frontmatter@RRAPformat{\produce@RRAP{*#1\href{mailto:#2}{#2}}}\frontmatter@RRAPformat}
  {}{}
}%
\makeatother

\usepackage[font=small,labelfont=bf,format=plain]{caption}

\usepackage{setspace}
  \usepackage{parskip}
\usepackage[compact]{titlesec}         
\titlespacing{\section}{4pt}{4pt}{4pt} 
\AtBeginDocument{
  \setlength\abovedisplayskip{4pt}
  \setlength\belowdisplayskip{4pt}
  }
  
\usepackage[]{caption, subfig}
\usepackage{lineno}
\setcitestyle{square}
\begin{document}

\newcommand{\ra}[1]{\renewcommand{\arraystretch}{#1}}
\newcommand{\deriv}{\mathrm{d}}

\title{Coupled 2-D MHD and runaway electron fluid simulations of SPARC disruptions}
\author{R. Datta}
\thanks{rdatta@mit.edu}
\affiliation{Plasma Science and Fusion Center, Massachusetts Institute of Technology, MA 02139, Cambridge, USA\looseness=-10000 
}%

\author{C. Clauser}
\affiliation{Plasma Science and Fusion Center, Massachusetts Institute of Technology, MA 02139, Cambridge, USA\looseness=-10000 
}%

\author{N. Ferraro}
\affiliation{Princeton Plasma Physics Laboratory, P.O. Box 451, Princeton, NJ 08543-0451, USA\looseness=-10000 
}%

\author{C. Liu}
\altaffiliation[Current address: ]{State Key Laboratory of Nuclear Physics and Technology, School of Physics, Peking University, Beijing 100871, China}
\affiliation{Princeton Plasma Physics Laboratory, P.O. Box 451, Princeton, NJ 08543-0451, USA\looseness=-10000 
}%

\author{R. Sweeney}
\affiliation{Commonwealth Fusion Systems, Devens, MA 01434, USA\looseness=-10000 
}%

\author{R. A. Tinguely}
\affiliation{Plasma Science and Fusion Center, Massachusetts Institute of Technology, MA 02139, Cambridge, USA\looseness=-10000 
}%



\begin{abstract}

Runaway electrons (REs) generated during disruption events in tokamaks can carry mega-Ampere level currents, potentially causing damage to plasma-facing components. Understanding RE evolution during disruption events is important for evaluating strategies to mitigate RE damage. Using two-dimensional toroidally symmetric magnetohydrodynamic (MHD) simulations in M3D-C1, which incorporates a fluid RE model evolved self-consistently with the bulk MHD fluid, we examine the seeding and avalanching of REs during disruptions in the SPARC tokamak – a compact, high-field, high-current device designed to achieve a fusion gain $Q > 2$ in deuterium–tritium plasmas. The M3D-C1 simulations of unmitigated disruptions demonstrate RE plateau formation and peaking of the final current density, which agree well with the results of lower-fidelity reduced RE fluid models. This work provides the first systematic comparison and benchmarking of different primary sources, including activated tritium beta decay and Compton scattering, in SPARC disruption simulations with self-consistent MHD and RE coupling.


\end{abstract}

\maketitle

\section{Introduction}
\label{sec:intro}

Runaway electrons (REs) are relativistic electron populations generated by strong electric fields during tokamak operation \citep{gill2002behaviour,granetz2014itpa,breizman2019physics}. During disruptions, which refer to the abrupt loss of plasma confinement typically initiated by large-scale plasma instabilities, RE beams with $\sim$MA-level currents can be generated \citep{gill2002behaviour,breizman2019physics,jepu2024overview}. The interaction of high-energy RE beams with plasma facing components can cause damage, often in the form of localized melting, vaporization, and brittle failure of these components \citep{jepu2019beryllium,jepu2024overview,ratynskaia2025modelling,ataeiseresht2023runaway}. This makes it important to accurately predict RE dynamics, and to develop strategies to mitigate RE impacts in high-field high-current devices.

The thermal quench (TQ) represents the first phase of a disruption. Rapid cooling of the plasma occurs during this phase due to enhanced heat transport caused by magnetohydrodynamic (MHD) activity or due to strong radiative losses caused by impurity injection \citep{breizman2019physics,reux2015runaway}. The TQ is followed by the current quench (CQ), during which the plasma current decays on a relatively slower inductive time scale \citep{martin2017formation}. The high post-TQ resistivity $\eta_\parallel$ of the plasma, combined with the large post-TQ current density $\bf{j_\parallel}$, produces a large parallel electric field $E_\parallel = \eta_\parallel j_\parallel$. If acceleration due to the electric field exceeds collisional damping, which typically occurs when the field exceeds a critical value $E_{CH}$, electrons become runaway and can be accelerated to relativistic energies \citep{dreicer1959electron,connor1975relativistic,breizman2019physics,martin2015avalanche}. 

Various mechanisms can seed runaway electrons during a disruption event for large values of $E_\parallel/E_{CH}$. Primary sources generate new runaway electrons from the bulk population. One such example is the Dreicer source, which diffusively brings new electrons into the runaway regime via small-angle collisions between bulk electrons \citep{dreicer1959electron}.  In fusion-relevant conditions, activated sources — namely Compton scattering and tritium beta decay — can become important \citep{breizman2019physics, martin2017formation,vallhagen2020runaway,ekmark_2025}.  The Compton source describes the scattering of $\sim$MeV-energy photons, generated from activated walls, by bulk plasma electrons. Electrons whose energies exceed a critical runaway energy $W_c$ post-scattering become new runaway electrons. Similarly, in deuterium-tritium (D-T) plasmas, the natural beta decay of tritium produces electrons, and the fraction of the beta distribution with energies $> W_c$ contributes to the runaway source \citep{martin2017formation}. Finally, the hot-tail mechanism can also be a significant source of primary electrons, in cases that involve the collisional energy equilibriation of hot and cold electron populations \citep{smith2008hot}. High-energy electrons experience a lower collision frequency, and thus cool slower than the bulk thermal population, providing a seed of `hot' electrons that can become runaways.

Once primary sources produce a runaway seed, exponential multiplication of the initial runaway seed can occur. This `avalanching' process generates secondary runaway electrons via large-angle collisional interactions between the seed runaway population and the bulk plasma electrons \citep{rosenbluth1997theory,breizman2019physics,martin2015avalanche}. The avalanching process is typically responsible for the fast time scales of exponential runaway electron growth observed in experimental and numerical studies \citep{gill2002behaviour,martin2015avalanche}. 
The rates of primary seeding and secondary avalanching depend on various parameters of the bulk plasma; therefore, the temporal evolution of these parameters during a disruption can affect the resulting dynamics of the CQ and RE evolution. 

In SPARC – a compact high-field, high-current tokamak designed to achieve a fusion gain $Q > 2$ \citep{creely2020overview} – both unmitigated and mitigated disruptions with RE generation may occur \citep{sweeney2020mhd}. Although SPARC is designed to be robust to disruptions and RE impacts, it will serve as a testbed for RE mitigation mechanisms for future fusion power plants. SPARC will be equipped with massive material injection (MMI) via massive gas or shattered pellets \citep{sweeney2020mhd,reux2015runaway,zhao2024simulation}, and the runaway electron mitigation coil (REMC) \citep{sweeney2020mhd,tinguely2023minimum,battey2023design}, which deconfines REs by generating field line stochasticity \citep{papp2013effect,papp2015energetic}. Predictions of the RE evolution in SPARC were performed previously by first modeling the disruption physics using the 3-D non-linear magnetohydrodynamic (MHD) code NIMROD \citep{glasser1999nimrod}, and by coupling the transport due to the observed MHD activity with the RE modeling framework DREAM \citep{hoppe2021dream,tinguely2021modeling,tinguely2023minimum}. In disruptions with neon injection and without any REMC effects, these simulations indicated a dominant contribution by the hot-tail seed ($I_{\text{seed}} \approx \SI{2}{\kilo\ampere}$), subsequently amplified via avalanching to produce a runaway plateau current of $I_f \approx \SI{5.5}{\mega \ampere}$ \citep{tinguely2021modeling}.


\begin{figure*}[t!]
\includegraphics[page=1,width=1.0\textwidth]{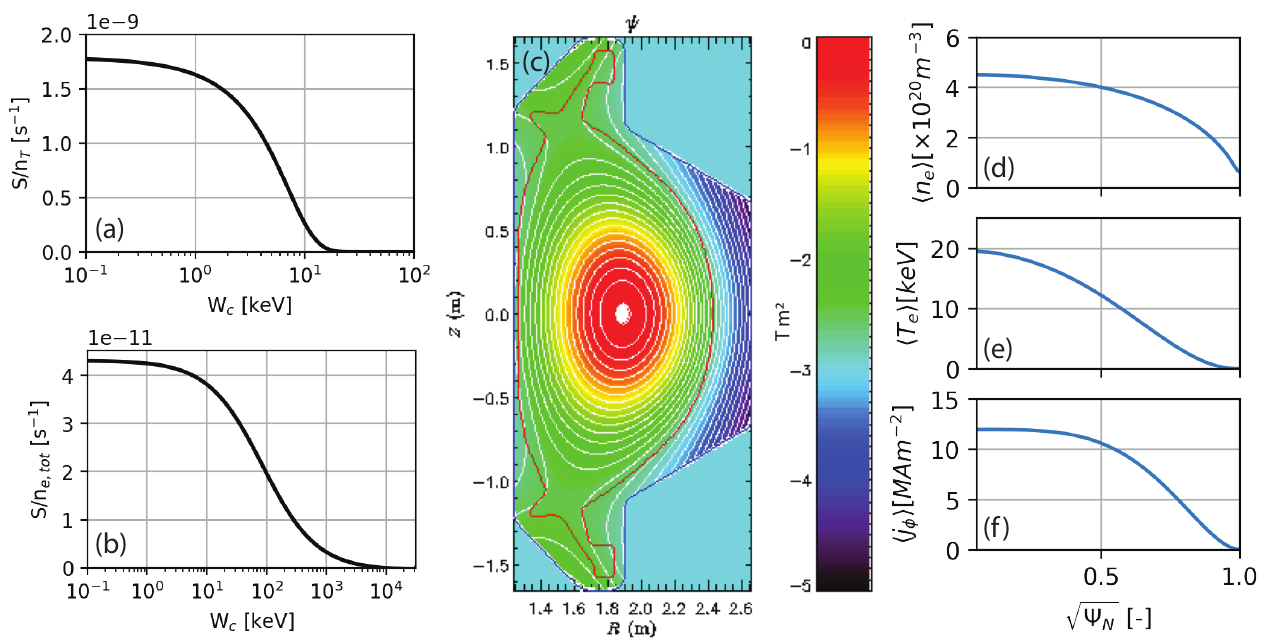}
\centering
\caption{ (a) Variation of the tritium beta decay rate with critical runaway energy $W_c$. (b) Variation of the Compton rate with critical runaway energy $W_c$, calculated using the estimated prompt $\gamma$-flux in the SPARC vacuum vessel. (c) Equilibrium distribution of the poloidal flux $[\SI{}{\tesla\meter\squared}]$ at $t = \SI{0}{\milli \second}$. White contours show lines of constant flux. (d-f) Variation of the flux-averaged electron density, electron temperature, and  toroidal current density as a function of the square root of the normalized poloidal flux $\psi_N$ at $t = \SI{0}{\milli\second}$.
}
\label{fig:init_conditions}
\end{figure*}

In this framework, however, there was no feedback between the RE evolution and the background MHD evolution. Recent work has explored such coupling between RE and MHD physics using codes, such as M3D \citep{cai2015influence}, M3D-C1 \citep{zhao2020simulation,zhao2024simulation} and JOREK \citep{bandaru2021magnetohydrodynamic,bandaru2024runaway}, which self-consistently evolve the runaway electron population and bulk MHD plasma. Such self-consistent modeling indicates that the coupling of REs with the bulk MHD fluid can produce important effects, such as modifying the growth and structure of MHD instabilities \citep{helander2007resistive,cai2015influence,zhao2020simulation,bandaru2021magnetohydrodynamic}, and reducing the runaway plateau in 3-D \citep{zhao2024simulation}. A coupled treatment is also important for evaluating  RE mitigation strategies, such as MMI and REMCs, which produce significant MHD activity. 

In this paper, we explore RE generation during unmitigated disruptions in SPARC, through coupled RE fluid and MHD simulations in a two-dimensional toroidally symmetric geometry, using the extended MHD code M3D-C1 \citep{jardin2012multiple}. This work provides a systematic comparison of various primary RE seeds in M3D-C1, including activated sources, relevant during unmitigated disruptions. We demonstrate that the type of RE seed affects not only the final plateau current, but also the shapes of the final current density and safety factor $q$. The Dreicer mechanism dominates over the tritium beta decay and Compton seeds in these unmitigated disruptions, demonstrating an off-axis peaking of the final current density and a hollow $q$-profile. In contrast, when the Dreicer term is artificially suppressed, the activated sources produce a much smaller RE plateau, and generate a more strongly peaked (on-axis) final current density profile with a low on-axis safety factor $q_0 < 1$. This work also benchmarks the M3D-C1 results against simpler RE fluid models \citep{martin2015avalanche,martin2017formation}, which demonstrate good agreement with the higher-fidelity M3D-C1 results, producing similar plateau currents, RE saturation times, and peaking of the final current density. Finally, we note that the simulations in this paper represent the first self-consistent RE and MHD simulations of SPARC disruptions, providing valuable insight into the relevance of various RE generation mechanisms in SPARC, and a basis for comparison of additional effects such as impurity injection and 3-D MHD activity.

The paper is structured as follows --- in \blue{\S }\ref{sec:methods}, we describe the simulation setup, initial and boundary conditions, and the relevant runaway electron physics incorporated into the M3D-C1 model. In \blue{\S }\ref{sec:results}, we describe post-thermal quench conditions, the resulting current quench, and the RE evolution. In particular, we compare results for a base case, where RE generation is artificially suppressed,  against cases where the REs are seeded via different primary mechanisms. \blue{\S }\ref{sec:discussion} provides analysis and discussion of key simulation results, and compares the M3D-C1 results with simpler reduced models. Finally, we reiterate key conclusions and outline future work in \blue{\S }\ref{sec:conclusions}.

\section{Methodology}
\label{sec:methods}

\subsection{Runaway Electron Fluid Model}

M3D-C1 is an extended MHD initial value code that employs high-order $C^1$ continuous finite elements in three dimensions \citep{jardin2012multiple}. In these simulations, we use a fluid runway electron model, the M3D-C1 implementation of which is described in \citeauthor{zhao2024simulation}. Here, we reiterate key aspects of the RE fluid model for the reader's convenience.  The coupling between the MHD and RE fluid equations is achieved primarily through the runaway current density ${\bf j_{RE}} = -en_{RE}{\bf v_{RE}} = {\bf j - j_{MHD}}$, where $e$ is the electron charge, $n_{RE}$ and ${\bf v_{RE}}$ are the runaway electron density and velocity, respectively, ${\bf j}$ is the total current density, and ${\bf j_{MHD}}$ is the current density carried by the bulk MHD fluid. The RE current density ${\bf j_{RE}}$ couples with the MHD equations via the momentum and energy equations, and through resistive Ohm's law \citep{zhao2020simulation,zhao2024simulation}. The MHD equations are closed by describing a set of fluid equations for the RE population, as shown below.

\begin{equation}
\begin{aligned}
& \partial_t n_{RE}+\nabla \cdot\left(n_{RE} \mathbf{v}_{RE}\right)=S_{RE} \\
& \mathbf{v}_{RE}=v_{\parallel} \mathbf{b}+\mathbf{v}_{E \times B} \\
& n_i=n_e+n_{RE}
\end{aligned}
\end{equation}

Here, $v_{\parallel}$ is the runaway electron velocity along the local magnetic field unit vector ${\bf b}$, ${\bf v}_{E \times B}$ is the ${\bf E \times B}$ drift velocity, and $n_i$ and $n_e$ are the ion and electron densities. $S_{RE} = S_\text{P} + \gamma_A n_{RE}$ represents the source term in the RE continuity equation, and has contributions from both primary $S_{P}$ and secondary avalanching $\gamma_A n_{RE}$.  The primary and secondary source terms, are in general, kinetic processes; however, we can approximate their contributions using analytical approximations. The (classical) Dreicer \citep{dreicer1959electron,connor1975relativistic} and Rosenbluth-Putvinski avalanching \citep{rosenbluth1997theory} sources were previously implemented in M3D-C1 \citep{zhao2024simulation}, and are described in \autoref{appen_A}. In the present work, we additionally consider the contributions of the tritium beta decay and Compton scattering sources, the implementation of which is described below. The hot-tail mechanism is not included in this work. 

We model the tritium beta decay source using \citep{martin2017formation}:

\begin{equation}
S_\beta = \ln 2 \frac{n_T}{\tau} \int_{W_c}^{W_\text{max}} f_\beta(E) dE
\label{eq:trit}
\end{equation}

Here, $n_T$ is the tritium density, $\tau \approx 4500$ days is the tritium half-life, $W_\text{max} = \SI{18.6}{\kilo\electronvolt}$ is the maximum energy of the $\beta$-electrons, $f_{\beta}(E)$ is the normalized $\beta$ energy spectrum \citep{stover}, and $W_c = m_ec^2[(1+1/E^{*})^{1/2}-1]$ is the critical runaway energy described earlier in \blue{\S }\ref{sec:intro}, defined in terms of the normalized electric field $E^* \equiv E_\parallel/E_{CH} = E_\parallel/ [e^3 n_e \ln \Lambda / (4 \pi \epsilon_0^2 m c^2)]$, where $E_{CH}$ is the critical Connor-Hastie electric field \citep{connor1975relativistic}. \autoref{fig:init_conditions}\blue{a} shows the variation of the tritium beta rate $S_\beta/n_T$ with critical runaway energy $W_c$. As $W_c \rightarrow 0$, the rate approaches a maximum value $S_\beta/n_T \approx \SI{1.8e-9}{\per \second}$. For $W_c > \SI{18.6}{\kilo\electronvolt}$, which represents the maximum energy of beta electrons, the rate drops to 0.

Similarly, the Compton source is given by \citep{martin2017formation}:

\begin{equation}
    S_C \approx n_{\mathrm{e}, tot} \int_{W_c} \Gamma_\gamma\left(E_\gamma\right) \sigma\left(E_\gamma\right) \mathrm{d} E_\gamma
\label{eq:compton}
\end{equation}

Where, $n_{e,tot}$ is the total electron density including free and bound contributions, $\Gamma_\gamma(E_\gamma)$ is the energy spectrum of the $\gamma$-flux in terms of the photon energy $E_\gamma$, and $\sigma(E_\gamma)$ is the Compton scattering cross-section, which can be obtained by integrating the Klein-Nishina differential cross-section over scattering angles greater than a critical value required to generate runaways (see Eq. 29 in \citeauthor{martin2017formation}). The prompt $\gamma$-flux $\Gamma(E_\gamma)$ inside the SPARC vacuum vessel (see \autoref{appen_B}) was obtained from OpenMC simulations of neutron and photon transport, assuming a fusion power of $\SI{140}{\mega\watt}$ \citep{romano2015openmc,Wang2025ANOpenMC}. The resulting Compton rate $S_C/n_{e,tot}$ is shown in \autoref{fig:init_conditions}\blue{b}. The maximum Compton rate, obtained for low values of $W_c$, is expected to be $S_C/n_{e,tot} \approx \SI{4e-11}{\per \second}$. We note that volumetric $\gamma$-generation from the bulk plasma, and geometric corrections to the $\gamma$-flux are not included in the present work.

\begin{figure*}[t!]
\includegraphics[page=2,width=1.0\textwidth]{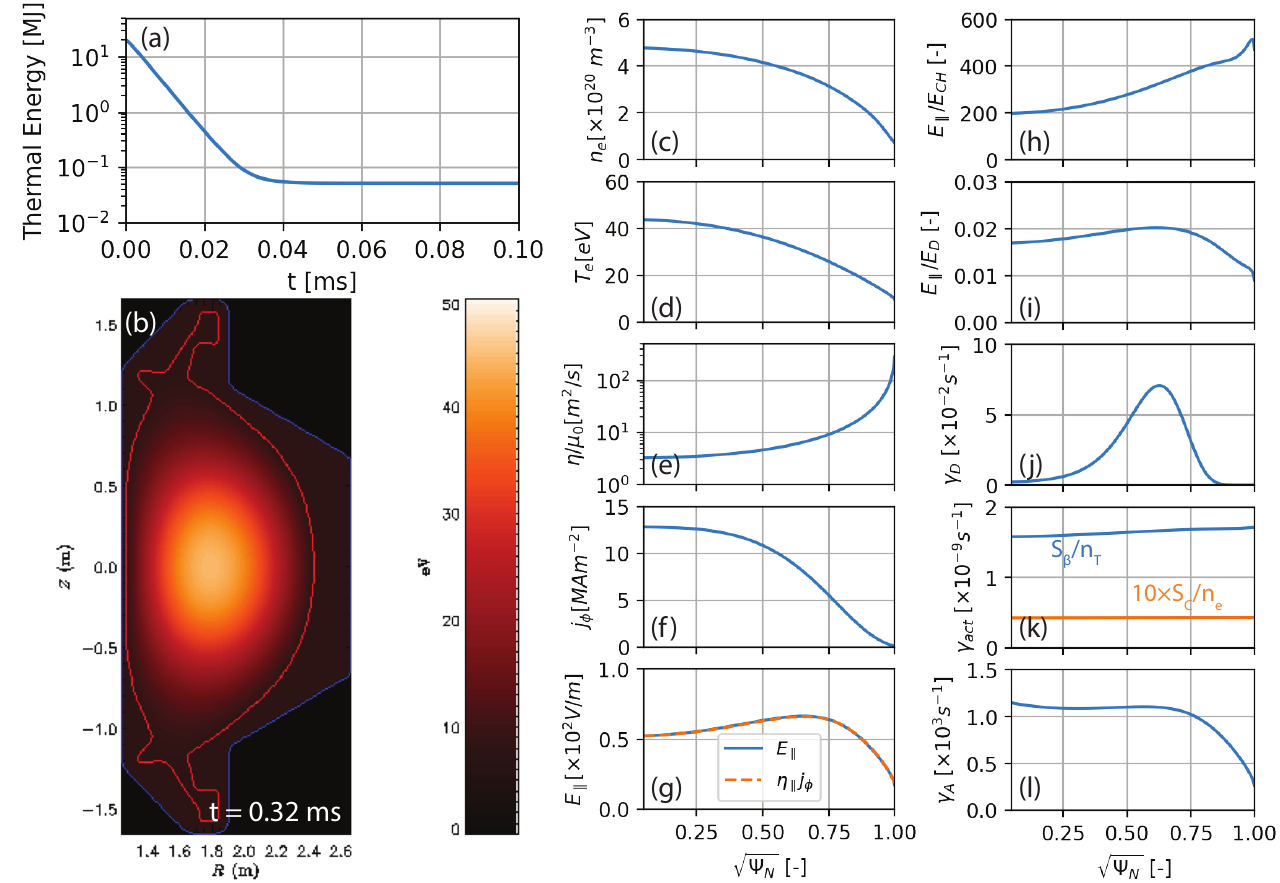}
\centering
\caption{ (a) Evolution of the total thermal energy during the thermal quench. The thermal energy decays rapidly due to thermal diffusion across the poloidal plane. (b) Distribution of the electron temperature in the poloidal plane at $t =\SI{0.32}{\milli\second}$. (c-g) Flux-averaged values of (c) electron density, (d) electron temperature, (e) resistive diffusivity, (f) toroidal plasma density, and (g) the parallel electric field at $t = \SI{0.32}{\milli\second}$. The parallel electric field at \SI{0.32}{\milli\second} normalized using the (h) Connor-Hastie critical field $E_{CH}$, and (i) the Dreicer field $E_D$. (j) The Dreicer source rate $S_D/n_e$, and (k) tritium beta $S_\beta/n_T$, and prompt Compton rate $S_C/n_e$ at $t = \SI{0.32}{\milli\second}$. We multiply the Compton rate by a factor of 10 for readability. (l) The Rosenbluth-Putvinski avalanching rate $\gamma_A$.}
\label{fig:TQ}
\end{figure*}

\subsection{Geometry and Initial Conditions}

We simulate disruptions of the primary reference discharge (PRD) in SPARC – which describes a D-T H-mode plasma with plasma current $I_p = \SI{8.7}{\mega\ampere}$, toroidal magnetic field $B_0 = \SI{12.2}{\tesla}$, and volume-averaged density and temperatures $\langle n_i \rangle \approx 3\times 10^{20} \SI{}{\per \meter \cubed}$ and $\langle T_e \rangle \approx \SI{7}{\kilo \electronvolt}$ \citep{creely2020overview,sweeney2020mhd,tinguely2021modeling}. 

The initial distribution of the equilibrium poloidal flux $\Psi (R,Z)$ is shown in \autoref{fig:init_conditions}\blue{c}, while the initial distributions of the flux-averaged electron density $n_e$, electron temperature $T_e$, and toroidal current density $j_\phi$ are shown in \autoref{fig:init_conditions}\blue{(d-f)}. As illustrated in \autoref{fig:init_conditions}\blue{c}, we use a simplified SPARC geometry, comprising a resistive first wall (in red), while the inner vessel (in blue) marks the boundary of the simulation domain. The minor and major radii are $a = \SI{0.57}{\meter}$ and $R_0 = \SI{1.85}{\meter}$, respectively. We use a perfectly conducting boundary condition at the inner vessel, and set the resistivity to $\eta \approx 3\times10^{-3}\SI{}{\ohm\meter}$ in the region between the first wall and the inner vessel. The high resistivity enables RE solutions for the resistive wall case \citep{martin2015avalanche}, and limits induced currents of large magnitude, given that the first wall and vacuum vessel do not provide a toroidally continuous conducting path. To examine the effect of wall conductivity, a limited number of simulations were also run with a perfectly conducting (and simplified) first wall. The simulation domain is descretized using an unstructured finite element mesh, and the MHD and RE fluid equations are solved in the plasma region enclosed within the first wall. The simulations are performed in 2-D with toroidal symmetry, which means that MHD modes with toroidal mode numbers $n > 0$ are not captured in our simulations (in contrast, previous NIMROD simulations capture modes with $n\geq 1$ \citep{tinguely2021modeling}). Such 3-D MHD activity can lead to RE losses by generating field line stochasticity \citep{rechester1978electron,tinguely2023minimum,papp2015energetic,paz2019kink}. Since such losses are suppressed in 2-D, the simulations in this paper represent the largest RE currents achievable for a given case. However, we note that magnetic islands formed via 3-D MHD activity can also trap and confine REs to some extent \citep{papp2015energetic}. 

We consider various cases, including simulations with (a)~no RE sources, (b)~a constant initial RE seed current, (c)~Dreicer source only, and (d)~activated sources only.  In each of these simulations, we use the same simulation geometry and an identical set of initial conditions (as shown in \autoref{fig:init_conditions}). In each case, a disruption is triggered by setting the perpendicular thermal conductivity to a large value $\chi_\perp \approx \SI{1.5e5}{\meter \squared \per \second}$. The RE sources are turned off during the TQ, such that there is no RE generation and growth during this phase.  The sources are then turned on post-TQ for $t > \SI{0.32}{\milli \second}$. 

We model the MHD fluid as a single species deuterium plasma, without any impurities. Although the plasma is modeled as a single deuterium species, the contribution of tritium beta decay to RE generation is accounted for by assuming that the tritium density is half the total ion density. We also note that impurities can affect the rates of the RE sources described in the previous subsection \citep{martin2015runaway,hesslow2019evaluation,hesslow2019influence}. For instance, partial screening due partially ionized impurities can significantly reduce the Dreicer rate \citep{hesslow2019evaluation}. The Compton rate can also be higher for large impurity fractions, since the Compton source scales with the total electron density $n_{e,tot}$ (\autoref{eq:compton}). Similarly, impurities can modify the critical runaway energy $W_c$ \citep{martin2015runaway}, and enhance the avalanching rate as a result of collisions between runaway and bound electrons \citep{hesslow2019influence}. Lastly, impurities not only provide  radiative cooling, they can also generate MHD instabilities, which can have important effects on the temporal evolution of the bulk plasma parameters, necessitating a 3-D treatment for accurate modeling \citep{ferraro20183d}. While the present work does not include these effects, our results provide an important understanding of RE dynamics in unmitigated disruptions, self-consistently simulating RE evolution with the MHD equations, and analyzing the relevance of different RE source terms in the SPARC tokamak. The effects of impurity injection and 3-D MHD activity  will be pursued in a future publication. 

\section{Results}
\label{sec:results}

\subsection{Thermal Quench}
\label{Sec:TQ}

\begin{figure}[t!]
\includegraphics[page=3,width=0.48\textwidth]{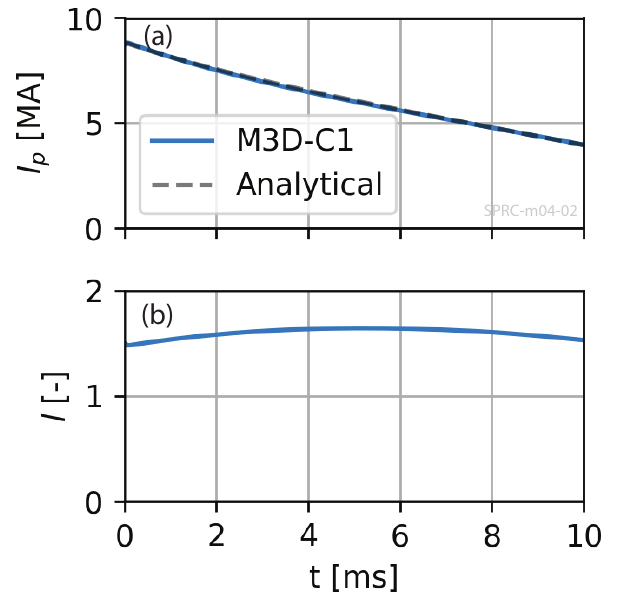}
\centering
\caption{ (a) Evolution of plasma current $I_p$ with time calculated by M3D-C1 for the simulation with no RE sources. The black dashed line represents the analytical solution for the current decay $I_p(t) =I_0 \exp(-t/\tau_{CQ})$, where $\tau_{CQ}$ is the inductive time of the plasma. (b) Evolution of normalized inductance $l$ with time, as calculated by M3D-C1.}
\label{fig:noRE}
\end{figure}

As mentioned in the previous section, we model the TQ identically in each simulation. The temporal evolution of the total thermal energy during the TQ is shown in \autoref{fig:TQ}\blue{a}. The thermal energy decays rapidly because of cross-field thermal diffusion, generated by a large value of the perpendicular thermal diffusivity $\chi_\perp$. The e-folding time of the exponential decay is roughly consistent with the thermal diffusion time $\tau_\text{diff.} \sim a^2/\chi_\perp \approx \SI{0.002}{\milli\second}$. Thermal equilibrium is re-established around $t_{TQ} \approx \SI{0.05}{\milli\second}$, and the thermal energy reaches a plateau of about $\SI{0.05}{\mega\joule}$. We note that this time is consistent with the shortest fast TQ durations ($\approx \SI{50}{\micro \second}$) estimated for SPARC \citep{sweeney2020mhd}. \autoref{fig:TQ}\blue{b} shows the distribution of the electron temperature $T_e$ in the poloidal plane at $t =\SI{0.32}{\milli\second}$, which is when the RE generation is turned on. By this time, the core temperature has fallen from about $T_e \approx \SI{20}{\kilo\electronvolt}$ initially to about $T_e \approx \SI{40}{\electronvolt}$. The post-TQ temperature is set primarily by the balance between thermal conduction and Ohmic heating.

\autoref{fig:TQ}\blue{(c-g)} show the post-TQ flux-averaged values of various plasma quantities at $t = \SI{0.32}{\milli\second}$. The large drop in the electron temperature causes a large plasma resistivity (see \autoref{fig:TQ}\blue{e}). In these simulations, we use the parallel Spitzer resistivity, which scales with temperature as $\eta_\parallel \sim T_e^{-3/2}$, causing the resistivity to rise by a factor of $>10^4$. The toroidal current density (\autoref{fig:TQ}\blue{f}) remains large, and similar to its value at $t =\SI{0}{\milli\second}$ (see \autoref{fig:init_conditions}\blue{f}). We note that an $I_p$ spike is not observed in these 2-D simulations. The combination of large plasma resistivity and high parallel current density $j_\parallel \approx j_\phi$ of the post-TQ plasma results in a large parallel electric field $E_\parallel = {\bf E \cdot b} = \eta_\parallel j_\parallel$ (\autoref{fig:TQ}\blue{g}). 

The post-TQ parallel electric field normalized in terms of the Connor-Hastie critical field $E_{CH} \propto n_e$ \citep{connor1975relativistic} and the Dreicer electric field $E_D \propto n_e/T_e$ \citep{dreicer1959electron} are shown in \autoref{fig:TQ}\blue{(h-i)}. The value of $E_\parallel/E_{CH}$ is large, increasing from roughly $E_\parallel/E_{CH} \approx 200$ near the core to about $E_\parallel/E_{CH} \approx 500$ near the edge. These large values of $E_\parallel/E_{CH} \gg 1$ are expected to produce REs through primary and secondary mechanisms. The Dreicer term is sensitive to  $E_\parallel/E_{D}$ (see \autoref{appen_A}), which has a maximum value of about $0.02$, corresponding to a Dreicer rate of $\gamma_D = S_D/n_e \approx \SI{0.07}{\per \second}$ as shown in \autoref{fig:TQ}\blue{j}. For the given values of $E_\parallel/E_{CH} \approx 200-500$, the critical runaway energy is $W_C<\SI{1}{\kilo\electronvolt}$. The resulting tritium beta and Compton rates (\autoref{fig:TQ}\blue{k}) are therefore roughly uniform and close to their maximum values, as seen from \autoref{fig:init_conditions}\blue{(a-b)}. However, these rates are much lower than the Dreicer rate observed in \autoref{fig:TQ}\blue{j}. Finally, \autoref{fig:TQ}\blue{l} shows the avalanching rate $\gamma_A$. The term $\tau^{-1}(E_\parallel/E_{CH}-1) \sim E_{CH}(1-E_\parallel/E_{CH})$ in \autoref{eq:ava} dominates the shape of the avalanching rate, while the factor $\zeta$ (varying between $\zeta \approx 0.4-1$ for SPARC) provides a slight geometric correction (see \autoref{appen_A}). The rate is high $\gamma_A \sim \SI{1e3}{\per \second}$ for $\sqrt{\psi_N}<0.75$, and then falls sharply near the edge. The next subsections describe how these RE generation rates in the post-TQ plasma affect the current quench dynamics.

\begin{figure}[b!]
\includegraphics[page=4,width=0.49\textwidth]{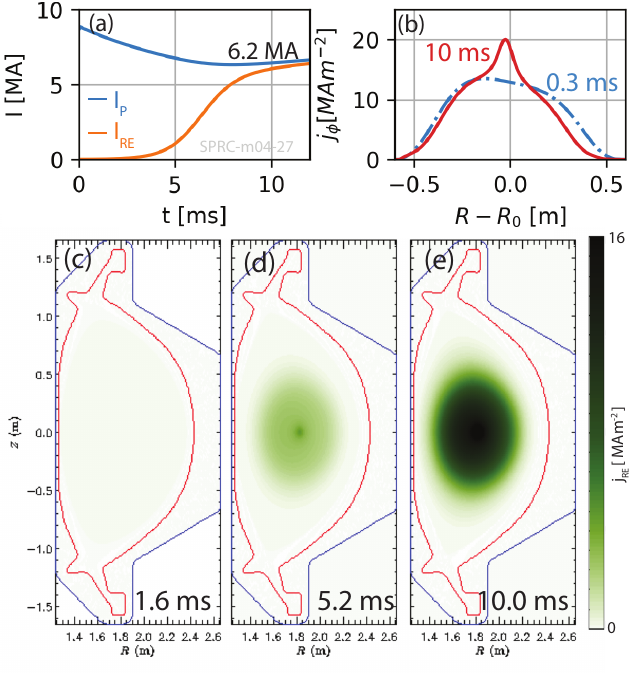}
\centering
\caption{ (a) Evolution of plasma current $I_p$ and RE current $I_{RE}$ for a disruption simulation with a $\SI{10}{\kilo\ampere}$ initial seed current and no other primary sources. (b) Torodial current density profile along $Z = \SI{0}{\meter}$ at \SI{10}{\milli\second} [red, solid] and \SI{0.3}{\milli\second} [blue, dashed]. Here, $R_0 = \SI{1.85}{\meter}$ is the major radius. (c-e) Distribution of the RE current density $j_{RE}$ in the poloidal plane at different times.}
\label{fig:initialSeed}
\end{figure}

\subsection{Current Quench and Runaway Electron Evolution}

\label{sec:CQ}

In the absence of any runaway electron sources, the plasma current simply decays with time, as shown in \autoref{fig:noRE}\blue{a}. The plasma current falls from an initial value of $I_p = \SI{8.7}{\mega \ampere}$ to about $\SI{4}{\mega \ampere}$ over $\SI{10}{\milli\second}$ since the start of the TQ. For a constant inductance $L \equiv 2W_\text{mag}/I_p^2$, where $W_\textbf{mag}$ is the total magnetic energy within the conducting inner vessel wall, we expect the plasma current to decay exponentially $I_p(t) =I_0 \exp(-t/\tau_{CQ})$, with a characteristic inductive time $\tau_{CQ} = L/\mathcal{R} = l\tau_\eta/4$ \citep{martin2015avalanche}. Here, $\mathcal{R}$ is the plasma resistance, $l = 2L/(\mu_0R_0)$ is the normalized inductance, and $\tau_\eta = a^2/(\eta/\mu_0)$ is the resistive time. \autoref{fig:noRE}\blue{b} demonstrates that the normalized inductance $l$ varies between $l \approx 1.5-1.6$ for $0 < t < \SI{10}{\milli\second}$. The dashed line in \autoref{fig:noRE}\blue{a} shows the analytical result, calculated using the mean value of the post-TQ resistivity (see \autoref{fig:TQ}\blue{e}) and the  inductance $L\approx \SI{1.8}{\micro\henry}$, demonstrating good agreement with the M3D-C1 result. 

\begin{figure}[t!]
\includegraphics[page=5,width=0.48\textwidth]{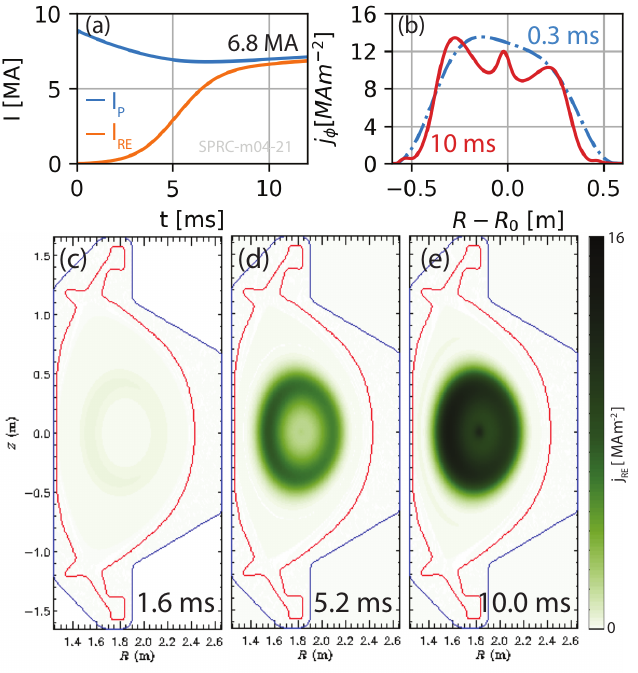}
\centering
\caption{ (a) Evolution of plasma current $I_p$ and RE current $I_{RE}$ for a disruption simulation with only the Dreicer primary source. (b) Torodial current density profile along $Z = \SI{0}{\meter}$ at \SI{10}{\milli\second} [red, solid] and \SI{0.3}{\milli\second} [blue, dashed]. Here, $R_0 = \SI{1.85}{\meter}$ is the major radius. (c-e) Distribution of the RE current density $j_{RE}$ in the poloidal plane at different times.}
\label{fig:dreicer}
\end{figure}

\autoref{fig:initialSeed} shows the M3D-C1 results for an initial RE seed current of $I_\text{seed} = \SI{10}{\kilo\ampere}$. The seed is applied by scaling the initial plasma current density shown in \autoref{fig:init_conditions}\blue{f}. All other primary sources were artificially suppressed in this simulation. The temporal evolution of the plasma current $I_p$ and RE current $I_{RE}$ are illustrated in \autoref{fig:initialSeed}\blue{a}. The plasma current $I_p$ falls from an initial value of $\approx \SI{8.7}{\mega\ampere}$ after the TQ, and the RE current $I_{RE}$ rises simultaneously. The current evolution saturates around $\SI{10}{\milli\second}$. The REs become the dominant current carrier, reaching a plateau of about $I_f\approx\SI{6}{\mega\ampere}$. \autoref{fig:initialSeed}\blue{(c-e)} show how the distribution of the RE current density $j_{RE}$ in the poloidal plane changes with time. The RE current density is high at the core and decreases towards the edge. This can be observed more clearly in \autoref{fig:initialSeed}\blue{b}, which shows the toroidal current density $j_\phi(R)$ along $Z = \SI{0}{\meter}$ at \SI
{10}{\milli\second}. Compared to the initial current density, the final current is more peaked and spans a smaller extent in the $R$-direction, resulting in a higher internal inductance $l_i \approx 1.4$, compared to the initial value $l_i (t = \SI{0}{\milli\second}) \approx 1.1$. The value of the internal inductance is calculated from the magnetic energy enclosed within the last closed flux surface. Finally, we note that although the current density near the core exceeds its initial value, the integrated value $I_p = \int j_\parallel(R,Z) dA_p$ over the poloidal area $A_p$ is lower at $t = \SI{10}{\milli\second}$.

\begin{figure}[b!]
\includegraphics[page=6,width=0.49\textwidth]{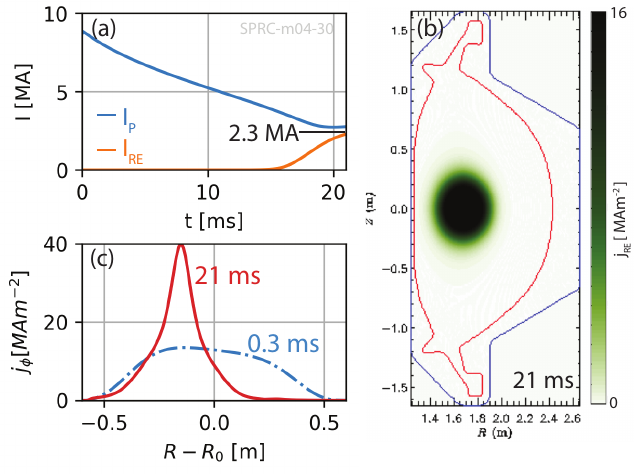}
\centering
\caption{ (a) Evolution of plasma current $I_p$ and RE current $I_{RE}$ for a disruption simulation with only the tritium beta and Compton sources. (b) Distribution of the RE current density $j_{RE}$ in the poloidal plane at \SI{21}{\milli\second}. (c) Torodial current density profile along $Z = \SI{0}{\meter}$ at \SI{21}{\milli\second} [red, solid] and \SI{0.3}{\milli\second} [blue, dashed]. Here, $R_0 = \SI{1.85}{\meter}$ is the major radius.}
\label{fig:activated}
\end{figure}

\autoref{fig:dreicer}\blue{a} shows the plasma and RE currents as a function of time for the simulation with only the Dreicer primary source. This simulation was initiated with a zero seed RE current, and all other primary sources were turned off. Similar to the previous case, the RE current rises and saturates at around \SI{10}{\milli\second}. The plateau current is $I_f\approx\SI{7}{\mega\ampere}$, higher than the case with the $\SI{10}{\kilo\ampere}$ initial seed. \autoref{fig:dreicer}\blue{(c-e)} show that the RE current density $j_{RE}$ distribution in the poloidal plane is inhomogeneous --- we observe a hollow density profile at $\SI{1.6}{\milli\second}$ (\autoref{fig:dreicer}\blue{c}) and $\SI{5.2}{\milli\second}$  (\autoref{fig:dreicer}\blue{d}), with a peak in $j_{RE}$ around $r \approx 0.2-\SI{0.3}{\meter}$. This is also apparent in \autoref{fig:dreicer}\blue{b}, which shows the toroidal current density $j_\phi(R)$ along $Z = \SI{0}{\meter}$ at \SI{10}{\milli\second}. In addition to a peak near the core, we observe an off-axis concentration of the current density. The generation of this current density profile will be discussed in more detail in \S\ref{sec:discussion}.

Finally, we present results for the case with only activated sources --- tritium beta and Compton --- in \autoref{fig:activated}. The tritium source is computed using \autoref{eq:trit}, assuming that half of the plasma fuel is composed of tritium, i.e. $n_T = 0.5n_i$. The Compton contribution is given by \autoref{eq:compton}; however, for these low post-TQ temperatures, we model the effect of non-prompt $\gamma$-flux, by multiplying the prompt flux by a factor of $10^{-3}$, similar to previous modeling efforts \citep{ekmark_2025}. As noted earlier, the rates of the activated sources are near their maximum values right after the TQ, and without any significant impurities, the activated rate is dominated by the tritium term. As observed in \autoref{fig:activated}\blue{a}, the RE current increases from a small value, because for the small magnitude of seed provided by the activated sources (as seen in \autoref{fig:TQ}\blue{k}), and is amplified via avalanching to eventually produce a plateau current of about $\SI{2}{\mega \ampere}$. The RE growth is much slower than the previous cases --- saturation occurs at around $\SI{20}{\milli\second}$, compared to around $\SI{10}{\milli\second}$ with the Dreicer source. The final distribution of the RE current density is shown in \autoref{fig:activated}\blue{b}. We observe high current density near the core; the current density spans a significantly smaller extent in the poloidal plane, and the magnetic axis has shifted towards the inboard side. This is further confirmed by \autoref{fig:activated}\blue{c}, which shows a highly peaked toroidal current density distribution, and significant contraction of the current density in the $R$-direction at $\SI{21}{\milli\second}$. 


\section{Discussion}
\label{sec:discussion}

\subsection{Discussion of M3D-C1 Results}


\begin{table}\centering
\ra{1.3}
\caption{
A summary of M3D-C1 simulation results, showing the plateau current $I_f$, final normalized internal inductance $l_i$, and the final safety factor on-axis $q(0)$. In each simulation, the initial values are $I_p \approx \SI{8.7}{\mega\ampere}$, $l_{i}\approx 1.1$,  and $q(0) \approx 0.93$.
}
\begin{tabular}{ccccc}
\hline
 Source & Wall & $I_f \, [\SI{}{\mega\ampere}]$ & Final $l_i$ & Final $q(0)$  \\
\hline
\SI{10}{\kilo\ampere} Seed & Resistive & 6.2 & 1.4 & 0.6 \\
Dreicer & Resistive & 6.8 & 1.3 & 1.0 \\
Dreicer ($2\times\eta_\parallel$) & Resistive & 7.8 & 1.2 & 1.0 \\
Activated & Resistive &  2.3 & 2.1 & 0.3 \\
Dreicer & Conducting & 5.3 & 1.6 & 1.0 \\
Activated & Conducting & 0.5 & 2.3 & 0.6 \\
\hline
\hline
\end{tabular}
\label{tab:table}
\end{table}

\autoref{tab:table} summarizes the M3D-C1 simulation results. The Dreicer seed dominates in the unmitigated disruptions, as seen both from the large rate of the Dreicer generation, compared to that of the activated sources, in the post-TQ plasma [see \autoref{fig:TQ}\blue{(j-k)}], as well as from the relatively higher RE current ($I_f \approx \SI{7}{\mega\ampere}$) and faster RE saturation observed in the Dreicer case (see \autoref{fig:dreicer}). An approximate value of the seed current can be estimated as the value when the the primary generation rate $S_P$ becomes comparable to the avalanching rate $\gamma_A n_{RE}$ \citep{martin2015avalanche}. This gives a seed current of roughly $I_\text{seed}\sim \SI{20}{\kilo\ampere}$ for the Dreicer case. In contrast, the estimated seed current for the activated case is approximately $I_\text{seed}\sim \SI{0.1}{\ampere}$. Activated sources generate a much smaller plateau current $I_f \approx \SI{2}{\mega\ampere}$, as shown in \autoref{fig:activated}, comparable to predictions made by other RE modeling frameworks such as DREAM for SPARC-like conditions \citep{ekmark_2025}. As noted earlier in \blue{\S }\ref{Sec:TQ}, the tritium beta source exceeds the Compton source by several orders of magnitude in the post-TQ plasma. However, as the RE current increases and the electric field $E_\parallel/E_{CH}$ consequently falls (and $W_c$ increases), the tritium beta source is suppressed first. This is because, as seen from \autoref{fig:init_conditions}\blue{a}, while the tritium beta source turns off completely for critical energies $W_C > \SI{18.6}{\kilo\electronvolt}$, the Compton contribution does not fall significantly until $W_c \gtrsim  \SI{100}{\kilo\electronvolt}$. However, since $E_\parallel/E_{CH}$ rises on the avalanching timescale, and given the much smaller Compton source in our case, the additional time that the Compton source is important does not significantly change the resulting RE dynamics.

\begin{figure}[t!]
\includegraphics[page=9,width=0.48\textwidth]{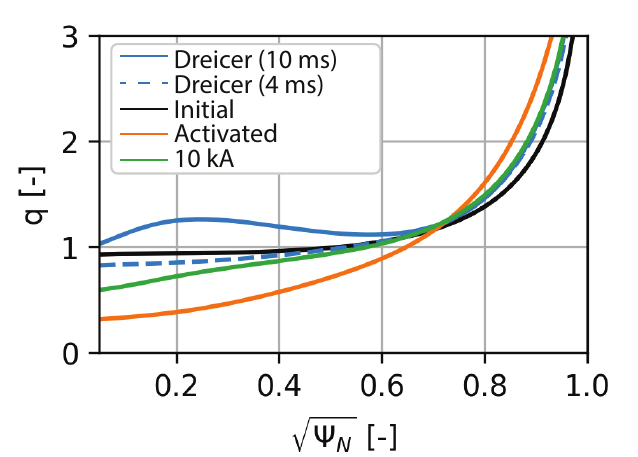}
\centering
\caption{ Safety factor $q$ as a function of the normalized poloidal flux computed by M3D-C1. For the Dreicer source (blue), we show $q$ at $\SI{4}{\milli\second}$ and at $\SI{10}{\milli\second}$. For the prescribed initial 10 kA seed (green), and the activated sources (orange), we show $q$ at $\SI{10}{\milli\second}$ and $\SI{21}{\milli\second}$ respectively.}
\label{fig:safety_factor}
\end{figure}

A similar argument also applies when $E_\parallel/E_{CH}$ rises rapidly during the TQ – the Compton seed is expected to become important first compared to the tritium seed. Furthermore, the contribution of prompt $\gamma$-s, which exceed that of the delayed (non-prompt) $\gamma$ flux, will be important at high temperatures. While we do not simulate RE generation during the TQ in our simulations, the relative contributions of the activated sources will likely vary with the TQ timescale. For instance, the maximum possible prompt Compton contribution (assuming $W_c \rightarrow 0$), for TQ times between $0.1-\SI{1}{\milli\second}$ and $n_e \sim \SI{1e20}{\per\cubic\meter}$, is approximately $n_C \sim S_C\tau_{TQ} \sim 10^5\text{-}10^6\,\SI{}{\per \cubic \meter}$. In reality, not all of these energetic electrons will runaway since the temperature must fall sufficiently for the electric field to exceed the critical value, but they can persist on the order of collisional timescales to generate a population of non-thermal energetic electrons that can be accelerated when $E_\parallel$ becomes greater than $E_{CH}$. The slowing down time for electron energies in the range $0.1-\SI{1}{\mega\electronvolt}$ is about $\sim 10^{-3}-10^{-1}\SI{}{\second}$ \citep{richardsonnrl2019}, indicating that some of the energetic electrons produced by Compton scattering of prompt $\gamma$-s may provide a delayed RE source after the TQ. This analysis, however, is beyond the scope of the current work, and likely requires a kinetic treatment of the non-thermal electron population.


The type of seed not only affects the RE plateau current, but also affects the shape of the final current density profile. In each of the simulations, we observe a on-axis peaking of the final current density $j_\phi$, resulting in an increase of the normalized internal inductance $l_i$ (see \autoref{tab:table}). Peaked current density profiles are characteristic of RE generation \citep{gill2002behaviour,eriksson2004current,martin2015avalanche,zhao2024simulation,tinguely2023minimum}, and are related to the radial diffusion of the electric field during avalanching \citep{eriksson2004current,smith2006runaway}. We observe that the peaking is highest for activated sources (\autoref{fig:activated}\blue{c}), where the initial RE seed is the smallest, reaching a final normalized inductance of $l_i \approx 2.1$. This is consistent with previous results that indicate a stronger peaking of the current profile for lower RE seeds \citep{martin2015avalanche}.


The peaking of the current profile can have important consequences for MHD stability. \autoref{fig:safety_factor} shows the final safety factor $q$ profile in each simulation. For the prescribed initial seed and the activated sources, the safety factor on-axis $q_0$ falls below 1. Low safety factors are associated with poor MHD stability – in 3-D, we may expect the excitation of internal kink, resistive kink, or tearing modes that modify the subsequent evolution of the plasma and RE dynamics \citep{helander2007resistive,cai2015influence,paz2019kink}. For the Dreicer source, the safety factor initially falls ($l_i$ is maximum around $\SI{4}{\milli\second}$ for the Dreicer simulation), as shown in \autoref{fig:safety_factor}. The final safety factor is $q > 1$ in this case, and demonstrates a hollow profile, which can be susceptible to tearing-type modes \citep{maget2007nonlinear}. The final on-axis safety factors achieved in each simulation are summarized in \autoref{tab:table}.


In disruptions with the prescribed initial seed and with the activated sources, which provide a mostly uniform initial RE seed, the final current density peaks near the core. In the disruption with the Dreicer seed, however, the current density exhibits an additional off-axis peak, as observed in \autoref{fig:dreicer}. As illustrated in \autoref{fig:TQ}\blue{g}, the parallel electric field is maximum off-axis near $\sqrt{\psi_N} \approx 0.7$, given increasing $\eta \sim T_e^{-3/2}$ and decreasing $j_\parallel$ with the minor radius $r$. The normalized field $E_\parallel/E_D$ is also maximum around this flux surface (\autoref{fig:TQ}\blue{i}), and so is the Dreicer rate (\autoref{fig:TQ}\blue{j}); however, the peak in the Dreicer rate is much sharper than that in $E_\parallel/E_D$, given the exponential dependence of the rate on $E_\parallel/E_D$. Consequently, primary electrons are preferentially seeded and eventually amplified near the  $\sqrt{\psi_N} \approx 0.7$ flux surface, resulting the the off-axis peaking of the RE current seen in \autoref{fig:dreicer}. 

\begin{figure}[t!]
\includegraphics[page=7,width=0.48\textwidth]{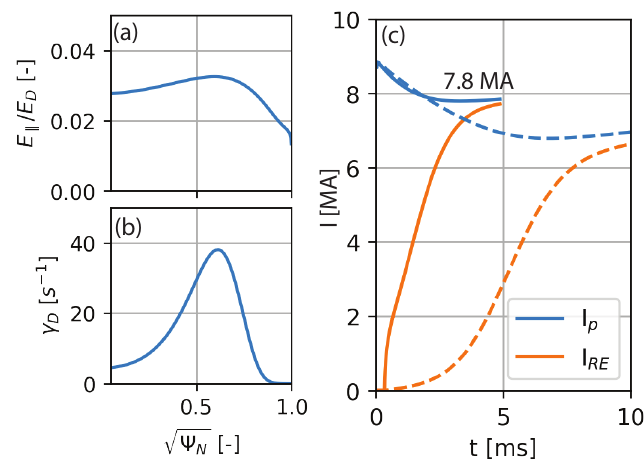}
\centering
\caption{ (a) The flux-averaged parallel electric field normalized by the Dreicer field $E_D$ at \SI{0.32}{\milli\second} for the higher resistivity case. (b) The flux-averaged Dreicer generation rate for the higher resistivity case. (c) The plasma and RE current evolution for the higher resistivity case with only the Dreicer source. The lower resistivity case (see \autoref{fig:dreicer}) is reproduced here (dashed lines) for comparison.}
\label{fig:high_resistivity}
\end{figure}

The exponential sensitivity of the Dreicer rate on $E_\parallel/E_D$ can have important consequences for the RE dynamics. To illustrate this, we repeat the simulation with Dreicer generation, but artificially increase the plasma resistivity $\eta$ by a factor of 2. \autoref{fig:high_resistivity} shows the results. The maximum value of $E_\parallel/E_D$ in the post-TQ plasma is about $0.03$, compared to about $0.02$ in the previous case. Note that $E_\parallel/E_D$ has not increased by $2\times$ because increased Ohmic heating $\eta|j|^2$ in the higher resistivity case raises the post-TQ equilibrium temperature. Despite the relatively small change in $E_\parallel/E_D$, the Dreicer rate increases by roughly 2 orders of magnitude, as shown in \autoref{fig:high_resistivity}\blue{b}. The larger seed therefore results in faster RE growth (saturation time $\approx \SI{4}{\milli\second}$), and a higher RE plateau ($I_f \approx \SI{8}{\mega \ampere}$), as seen in \autoref{fig:high_resistivity}\blue{c}.

\begin{figure}[b!]
\includegraphics[page=13,width=0.48\textwidth]{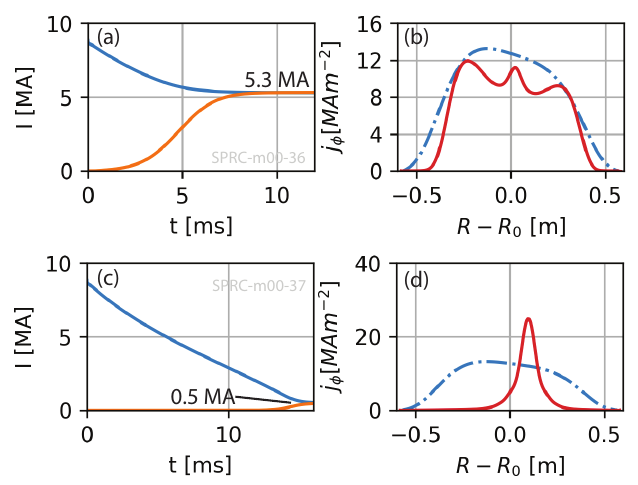}
\centering
\caption{ Results of M3D-C1 simulations with a perfectly conducting first wall. (a) The plasma and RE current evolution for Dreicer only case. The plateau current is lower compared to the resistive wall case, and the final current remains constant in time after plateau formation. (b) Initial [blue, dashed] and final [red, solid] profiles of torodial current density along $Z = \SI{0}{\meter}$. (c-d) Current evolution, and initial [blue, dashed] and final [red, solid] profiles of torodial current density along $Z = \SI{0}{\meter}$ for the activated sources only case.}
\label{fig:conductive}
\end{figure}

In each of the simulations, we also observe that the current continues to rise slowly after the initial plateau formation, which is characteristic of external magnetic field diffusion into the plasma volume through the resistive wall \citep{martin2015avalanche}. Indeed, despite the large resistivity of the first wall-inner vessel region, wall currents on the order of $\sim 10-\SI{20}{\kilo\ampere}$ are generated during the CQ in our simulations. To examine the impact of wall conductivity, we repeat the simulations with a perfectly conducting first wall to prevent any coupling with external magnetic fields. The results are shown in \autoref{fig:conductive}. As expected, these simulations demonstrate a lower plateau current since magnetic energy outside the conducting first wall is unavailable for runaway conversion, and the final current remains constant in time after plateau formation. For the Dreicer case,
the plateau current with the conducting first wall is $I_f \approx \SI{5.3}{\mega\ampere}$, while that for the activated case is about $I_f \approx \SI{0.5}{\mega\ampere}$. Consistent with the resistive first wall simulations, we observe an off-axis peaking of the final current density in the Dreicer case, while strong on-axis peaking is observed for the activated case.

\subsection{Comparison with Reduced RE Fluid Models}

 \begin{figure*}[t!]
\includegraphics[page=11,width=1.0\textwidth]{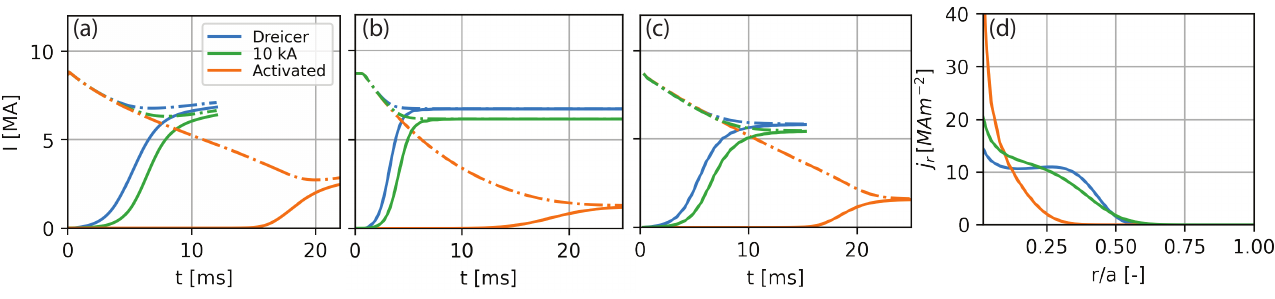}
\centering
\caption{(a) M3D-C1 predictions of the plasma and RE currents. Reproduced from \blue{\S}\ref{sec:results}. (b) Predictions of RE current $I_{RE}$ from the reduced 0-D (\autoref{eq:0-D}) for SPARC-like conditions. Dashed lines show the plasma current $I_p$, while solid lines represent the RE current $I_{RE}$. (c)~Predictions of RE current $I_{RE}$ from the reduced 1-D model (\autoref{eq:1D}).  (d) 1-D predictions of the final RE current density $j_{RE}$ as function of the normalized minor radius $r/a$, calculated using \autoref{eq:1D}.}
\label{fig:0D}
\end{figure*}


To gain further insight into the M3D-C1 results, we compare them to simpler reduced models of RE generation. Such models have been used extensively in the literature to predict and understand the behavior of REs in different tokamaks \citep{gill2002behaviour,eriksson2004current,martin2015avalanche,martin2017formation,fulop2020effect}. First, we consider a 0-D model, which is obtained by integrating the induction equation over the poloidal area $A_p$, and expressing the magnetic flux $\Phi \equiv LI_p$ in terms of the total inductance $L$ and total plasma current $I_p$ \citep{martin2015avalanche}:

\begin{equation}
\begin{split}
    \deriv_t I_p &\approx -2 \pi R_0 E_\parallel / L \\
    E_\parallel &= \eta_\parallel(I_p - I_{RE})/(\pi a^2) \\
    \deriv_t I_{RE}&=\left(\deriv_t I_{RE}\right)_{\text {primary }}+\left(\deriv_t I_{RE}\right)_{\text {avalanche }} \\
\end{split}
\label{eq:0-D}
\end{equation}

 \autoref{eq:0-D} represents an initial value problem with two coupled ordinary differential equations, where the coupling is achieved via the parallel electric field. We have additionally assumed a constant inductance $L$, and neglected any RE losses and coupling with the vessel. Furthermore, the 0-D model does not include any dependence on the spatial profile of the relevant plasma quantities. Nevertheless, this model can be used to gain qualitative insight into the relative importance of primary sources and ensuing RE evolution for SPARC-like conditions. In the absence of any RE current $I_{RE}$, \autoref{eq:0-D} reduces to $d_t I_p \approx -2 \pi R_0 \eta_\parallel I_p/(L \pi a^2)$, which can be solved exactly, giving the exponential decay $I_p(t) =I_0 \exp(-t/\tau_{CQ})$, shown earlier to be a good match to the M3D-C1 result in \autoref{fig:noRE}\blue{a}.

\autoref{fig:0D}\blue{b} shows the solutions of the reduced 0-D model for SPARC-like conditions. In each case, we use a density $n_i = \SI{3e20}{\per \meter \cubed}$, equal to the volume-averaged ion density for the SPARC PRD case, and inductance $L \approx \SI{1.8}{\micro\henry}$, as calculated by M3D-C1 for the initial equilibrium shown in \autoref{fig:init_conditions}. The TQ is prescribed as $T_{\mathrm{e}}(t)=T_{\mathrm{f}}+\left[T_0-T_{\mathrm{f}}\right] \mathrm{e}^{-t / \tau_{TQ}}$, using representative values of initial $T_0 = \SI{20}{\kilo\electronvolt}$ and final temperatures $T_f = \SI{10}{\electronvolt}$, and the RE seed and avalanching terms are calculated using the equations shown in \blue{\S }\ref{sec:methods}. Using these values, we roughly match the maximum value of $E_\parallel/E_D \approx 0.02$ observed in the M3D-C1 result (\autoref{fig:TQ}\blue{i}). The 0-D results largely agree with the higher-fidelity M3D-C1 simulations (reproduced in \autoref{fig:0D}\blue{a}), producing similar RE plateau currents $I_{RE}$ currents in each case. As observed in \autoref{fig:0D}\blue{b}, the Dreicer source produces the largest RE current and the fastest RE evolution, followed by the $\SI{10}{\kilo\ampere}$ initial seed case, which shows a slightly lower plateau, similar to the M3D-C1 result. The activated sources generate the lowest RE plateau and demonstrate the slowest RE dynamics, again consistent with the M3D-C1 results (\autoref{fig:activated}). When all sources are turned on, including the $\SI{10}{\kilo\ampere}$ initial seed, the 0-D result is identical to the Dreicer only case, demonstrating the dominant effect of the Dreicer term for the given conditions. 

As noted earlier, the 0-D model assumes a constant inductance $L$, which does not capture changes to the spatial distribution of the current density with time. A 1-D description can help incorporate this effect. This model can be obtained by taking the curl of the induction equation, and computing the component along the local magnetic field direction ${\bf b}$ \citep{martin2015avalanche}:

\begin{equation}
    \begin{split}
        \mu_0 \frac{\partial j_p}{\partial t}&=\frac{1}{r} \frac{\partial}{\partial r}\left[r \frac{\partial E_{\parallel}}{\partial r}\right] \\
        \frac{\partial j_{RE}}{\partial t}&=\left(\frac{\deriv j_{RE}}{\deriv t}\right)_{\text {primary }}+\left(\frac{\deriv j_{RE}}{\deriv t}\right)_{\text {avalanche }} \\
        E_{\|}&=\eta_{\|}\left(j_p-j_{RE}\right)
    \end{split}
    \label{eq:1D}
\end{equation}

Here, $j_p$ and $j_{RE}$ are the total plasma current density and the RE current density respectively, and $r$ is the minor radius. The 1-D model involves two partial differential equations, coupled using the electric field $E_\parallel$. We solve \autoref{eq:1D}, along with the symmetry $\partial_r j_{RE}(r=0)=\partial_r j_p(r=0)=0$ and conducting wall $E_{\|}(r=a)=0$ boundary conditions, using an explicit finite difference scheme. We use the flux-averaged quantities in the post-TQ plasma calculated by M3D-C1 [\autoref{fig:TQ}\blue{(c-g)}] directly as initial conditions for the problem. The electron density changes to maintain quasi-neutrality $n_i = n_e + n_{RE}$, consistent with the M3D-C1 implementation, and as noted earlier in \blue{\S }\ref{sec:results}, since the thermal diffusion time is much shorter than the CQ time, we determine the temperature from the steady-state balance between thermal conduction and Ohmic heating.  
We note that in addition to the perfectly conducting wall, this model also assumes no RE losses and a cylindrical geometry. Although the 1-D model can be extended to include additional effects, such as a resistive wall \citep{eriksson2004current,martin2015avalanche,papp2013effect}, plasma elongation \citep{fulop2020effect}, or RE losses \citep{papp2013effect,martin2015avalanche}, in this current work we neglect these corrections for simplicity.

The 1-D predictions for the plasma and runaway current evolution are shown in \autoref{fig:0D}\blue{c}. The 1-D results reproduce both the plateau currents and the RE growth time scales well. The plateau currents are slightly lower than in the M3D-C1 simulations described in \blue{\S }\ref{sec:results} (and reproduced in \autoref{fig:0D}\blue{a}), which is expected given the conductive wall boundary condition imposed in the 1-D reduced model. We show the final runaway current density profiles $j_{RE}$, as predicted by the 1-D model, in \autoref{fig:0D}\blue{d}. In each case, the final current density exhibits comparable magnitudes and a peaked structure, consistent with the M3D-C1 results. The 1-D model also reproduces the off-axis peaking of the current density in the Dreicer case (\autoref{fig:dreicer}\blue{b}), and the strong on-axis peaking observed in the activated case, consistent with the M3D-C1 result (\autoref{fig:activated}\blue{c}). 

The good agreement observed between the higher fidelity M3D-C1 results and the reduced 0-D/1-D models not only enables benchmarking of our simulation results, but also provides fast efficient models to examine RE evolution under simplified conditions. While the 0-D model can provide fast predictions of the plateau current for different primary seeds, the 1-D model can additionally predict changes to the current density profile, given initial profiles of the relevant plasma quantities.

\section{Conclusions}
\label{sec:conclusions}

We perform 2-D coupled magnetohydrodynamic (MHD) and runaway electron (RE) fluid simulations of unmitigated disruption events in the SPARC tokamak. We reiterate that although SPARC may experience some unmitigated disruptions, massive material injection (MMI) will be routinely used, where possible, for disruption mitigation \citep{sweeney2020mhd}. Our simulation results are compared for various cases, including with no RE sources, a constant initial RE seed current, Dreicer source, and finally, with activated tritium beta decay and Compton scattering sources. In each case, we observe RE growth and eventual saturation, with a peaking of the final current density profile, consistent with RE beam formation. For the given conditions, the Dreicer term dominates over the other sources, producing the fastest RE growth and eventual plateau of $I_f \approx \SI{7}{\mega\ampere}$ (\autoref{fig:dreicer}\blue{a}).  The Dreicer simulation also demonstrates strong off-axis peaking of the final RE current (\autoref{fig:dreicer}\blue{b}), which is consistent with the off-axis maximum in the parallel electric field, and the strong exponential dependence of the Dreicer term on the normalized field $E_\parallel/E_D$. When the Dreicer term is artificially suppressed, RE seeding by the activated sources produces a much smaller $I_f \approx \SI{2}{\mega\ampere}$ plateau current (\autoref{fig:activated}\blue{a}). The final current density exhibits much stronger peaking relative to the other cases, consistent with previous work that shows stronger peaking with smaller RE seeds (\autoref{fig:activated}\blue{b}). The M3D-C1 results are in good agreement with simpler reduced models of RE evolution in SPARC-like conditions, demonstrating similar plateau currents, RE growth time-scales, and peaking of the final current density profile (\autoref{fig:0D}), providing confidence for more complex simulations in upcoming publications. 

The coupled MHD and RE fluid simulations outlined in this work complement previous predictions of RE generation in SPARC \citep{sweeney2020mhd,tinguely2021modeling,tinguely2023minimum,ekmark_2025}, and provide insight into the various contributions of different RE source terms in high-field high-current tokamaks. Future work will extend the current methodology to incorporate additional effects, such as 3-D MHD activity and impurity injection. Due to the strong peaking and the resulting low safety factors observed in our simulations (see \autoref{fig:safety_factor}), we can expect the development of MHD instabilities which may interact with the RE fluid and modify the eventual evolution of the plasma and RE dynamics \citep{cai2015influence,paz2019kink,zhao2020simulation}. As noted earlier, MHD activity also generates magnetic field line stochasticity, producing runaway electron losses \citep{papp2013effect,papp2015energetic}. The self-consistent coupling of 3-D MHD activity and RE formation is important for evaluating RE mitigation strategies, such as RE mitigation coils \citep{tinguely2021modeling,battey2023design} and MMI, as described earlier in \blue{\S} \ref{sec:intro}. Efforts are currently underway to incorporate the effects of impurity injection, 3-D MHD activity, and REMC magnetic fields into the M3D-C1 simulations.

\section{Data Availability}

The data that support the findings of this study are available upon reasonable request from the authors.

\section{Declaration of Interests}

The authors have no conflicts of interest to disclose.

\section{Acknowledgments}

This work is supported by Commonwealth Fusion Systems. The simulations presented in this paper were performed on the the Engaging cluster at the MGHPCC facility (www.mghpcc.org). This research also used resources of the National Energy Research Scientific Computing Center (NERSC) [10.13039/100017223], a Department of Energy Office of Science User Facility (m3195 for year 2024). The authors acknowledge Jin Chen (PPPL) and Ellen Seegyoung Seol (RPI) for installing M3D-C1 in the MIT Engaging cluster and for their continued support. The authors also acknowledge X. Wang for generating the $\gamma$-flux using OpenMC, and A. Carter for generating the $\gamma$-flux using MCNP.

\begin{appendices}

\section{Appendix A}
\label{appen_A}

The Dreicer source $S_D$ in M3D-C1 is given by \citep{dreicer1959electron,breizman2019physics}:

\begin{equation}
S_D=n_e \nu_{e e} x^{-\frac{3}{16}\left(Z_{\text {eff }}+1\right)} \exp \left(-\frac{1}{4x}-\left[\frac{\left(Z_{\text {eff }}+1\right)}{x}\right]^{1 / 2}\right)
\end{equation}

Here, $\nu_{ee} = n_e e^4 \ln \Lambda /\left(4 \pi \epsilon_0^2 m_e^2 v_{t e}^3\right)$ is the electron-electron collision frequency, $x \equiv E_\parallel/E_D$ is the ratio of the parallel field to the Dreicer electric field $E_D = e^3 n_e \ln \Lambda /(4 \pi \epsilon_0^2 T_e) = E_{CH} m_e c^2/T_e$, $Z_\text{eff}$ is the effective ionization, $v_{te} = \sqrt{2T_e/m_e}$ is the electron thermal velocity, $\ln \Lambda$ is the Coulomb logarithm for relativistic electrons, $T_e$ is the electron temperature, and $E_{CH}$ is the critical Connor-Hastie electric field \citep{connor1975relativistic}. 

The Rosenbluth-Putvinski model \citep{rosenbluth1997theory} is used to describe the avalanching of secondary runaways $S_A = n_{RE}\gamma_A$:

\begin{equation}
\begin{aligned}
&\gamma_{A} =  \frac{1}{\tau \ln \Lambda} \sqrt{\frac{\pi \zeta}{3(Z_\text{eff}+5)}}\left(\frac{E_\parallel}{E_{CH}}-1\right) \\
& \times\left(1-\frac{E_{CH}}{E_\parallel}+\frac{4 \pi(Z_\text{eff}+1)^2}{3 \zeta(Z_\text{eff}+5)\left(E_\parallel^2 / E_{CH}^2+4 / \zeta^2-1\right)}\right)^{-\frac{1}{2}}
\end{aligned}
\label{eq:ava}
\end{equation}

Here, $\tau = m_ec/(eE_{CH})$, $\zeta \approx \left(1+1.46\sqrt{\epsilon} + 1.72\epsilon\right)^{-1}$, and $\epsilon=r/R$ is the inverse aspect ratio in terms of the minor $r$ and major $R$ radii .

\section{Appendix B}
\label{appen_B}

\begin{figure}[t!]
\includegraphics[page=10,width=0.48\textwidth]{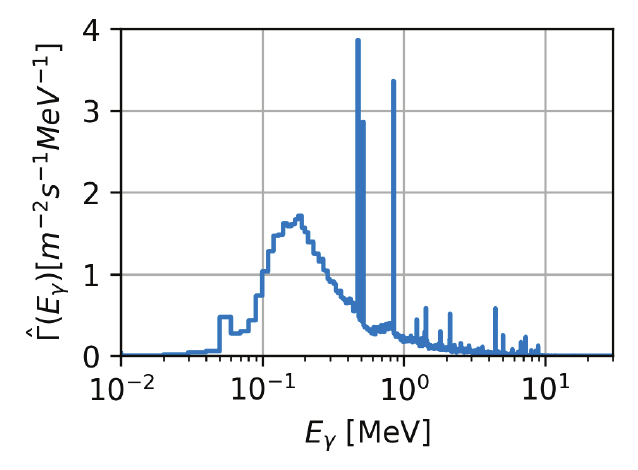}
\centering
\caption{The prompt $\gamma$-flux inside the vacuum vessel normalized by the total estimated flux $\Gamma_0 \approx \SI{1.6e18}{\per \meter \squared \per \second}$.}
\label{fig:gamma_flux}
\end{figure}

The prompt $\gamma$-flux $\Gamma(E_\gamma)$ inside the SPARC vacuum vessel from the activated walls, which is used to calculate the Compton rate (see \autoref{eq:compton}), was obtained from OpenMC simulations of neutron and photon transport, for the SPARC PRD case, assuming a fusion power of $\SI{140}{\mega\watt}$ and with uniform bin widths of \SI{10}{\kilo\electronvolt} \citep{romano2015openmc,Wang2025ANOpenMC}.
The prompt $\gamma$-flux inside the vacuum vessel normalized by the total estimated flux $\Gamma_0 \approx \SI{1.6e18}{\per \meter \squared \per \second}$ is shown in \autoref{fig:gamma_flux}. The prompt $\gamma$-flux in SPARC was additionally calculated using MCNP (Monte Carlo N-Particle), which agrees with the OpenMC result in terms of the total flux $\Gamma_0 \approx \SI{1.6e18}{\per \meter \squared \per \second}$, as well as the prompt $\gamma$-rate in \autoref{fig:init_conditions}\blue{b}. 

\newpage

\end{appendices}

\setcitestyle{square}
\bibliography{main}
\setcitestyle{square}

\end{document}